\documentclass[12pt]{article}
\usepackage{amssymb}
\usepackage{amscd}                                      

\renewcommand{\baselinestretch}{2.0}
\renewcommand{\baselinestretch}{1.17}%
\def\car{\mathop{\square}}
\def\carre#1#2{\raise 2pt\hbox{$\scriptstyle #1$}\car_{#2}}

\newcommand{\nc}{\newcommand}
\newcommand{\rnc}{\renewcommand}
\nc{\be}{\begin{equation}}
\nc{\ee}{\end{equation}}
\nc{\bea}{\begin{eqnarray}}
\nc{\eea}{\end{eqnarray}}

\newcommand {\newsection}{\setcounter{equation}{0}\section}

\def\slash#1{\setbox0=\hbox{$#1$}#1\hskip-\wd0\hbox to\wd0{\hss\sl/\/\hss}}

\def\href#1#2{{#2}}

\rnc{\a}{\alpha}
\nc{\ab}{\bar{\a}}
\nc{\adg}{a^{\dagger}}
\nc{\ap}{\a^{+}}
\nc{\abm}{\ab^{-}}
\rnc{\b}{\beta}
\nc{\bb}{\bar{\b}}
\nc{\bbp}{\bb_{\zb}^{+}}
\nc{\bm}{\b_{z}^{-}}
\nc{\oa}{\overline{\a}}
\nc{\ob}{\overline{\b}}
\rnc{\gg}{\gamma}
\rnc{\d}{\delta}
\nc{\f}{\phi}
\nc{\fb}{\bar{\phi}}
\nc{\vf}{\varphi}
\nc{\p}{\psi}

\rnc{\c}{\chi}
\nc{\la}{\lambda}
\nc{\m}{{\mathrm m}}
\nc{\n}{\nu}
\rnc{\o}{\omega}
\nc{\Om}{\Omega}
\rnc{\t}{\theta}
\nc{\eps}{\epsilon}
\rnc{\S}{\Sigma}
\nc{\F}{\Phi}
\nc{\trac}[2]{{\textstyle\frac{#1}{#2}}}
\nc{\ex}[1]{\mbox{e}^{\,\textstyle#1}}
\nc{\mat}[4]{\left(\begin{array}{cc}#1&#2\\#3&#4\end{array}\right)}
\nc{\som}[9]{\left(\begin{array}{ccc}#1&#2&#3\\#4&#5&#6\\#7&#8&#9%
\end{array}\right)}
\nc{\tr}{\mathop{\mbox{tr}}\nolimits}
\nc{\ad}{\mathop{\mbox{ad}}\nolimits}
\nc{\Tr}{\mathop{\mbox{Tr}}\nolimits}
\nc{\Det}{\mathop{\mbox{Det}}\nolimits}
\nc{\rk}{\mathop{\mbox{rk}}\nolimits}
\nc{\ra}{\rightarrow}
\nc{\Ra}{\Rightarrow}
\nc{\LRa}{\Leftrightarrow}
\nc{\ot}{\otimes}
\rnc{\ss}{\subset}
\nc{\nul}{\noindent\underline}
\nc{\non}{\nonumber\\}
\nc{\ld}{\overleftarrow}
\nc{\rd}{\overrightarrow}
\nc{\zb}{\bar{z}}
\rnc{\lg}{\frak{g}}
\nc{\lt}{\frak{t}}
\nc{\lk}{\frak{k}}
\nc{\lh}{\frak{h}}
\nc{\pik}{\Pi_{\lk}}
\nc{\pip}{\Pi_{+}}
\nc{\pim}{\Pi_{-}}
\nc{\pih}{\Pi_{\lh}}
\nc{\jz}{J_{z}}
\nc{\jzh}{\jz^{\lh}}
\nc{\jzp}{\jz^{+}}
\nc{\jzm}{\jz^{-}}
\nc{\del}{\partial}
\nc{\dz}{\del_{z}}
\nc{\dzb}{\del_{\bar{z}}}
\nc{\az}{A_{z}}
\nc{\azb}{A_{\bar{z}}}
\nc{\g}{g^{-1}}
\nc{\dw}{\Delta_{W}}
\nc{\Ad}{{\mbox{Ad}}}
\nc{\ks}{Ka\-za\-ma-\-Su\-zu\-ki}
\nc{\KS}{\ks}
\nc{\ksm}{\ks\ model}
\rnc{\AA}{{\mathbb A}}
\nc{\BB}{{\mathbb B}}
\nc{\CC}{{\mathbb C}}
\nc{\PP}{{\mathbb P}}
\nc{\rr}{{\mathbb {R}}}
\nc{\rrlam}{{\mathbb {R}}^3_{\lambda}}
\nc{\rrTth}{{\mathbb {R}}^2_{\theta}}
\nc{\rrTHth}{{\mathbb {R}}^3_{\theta}}
\nc{\zz}{{\mathbb {Z}}}
\nc{\cpm}{\CC\PP(m)}
\nc{\cpn}{\CC\PP(n)}
\nc{\cp}[1]{\CC\PP(#1)}
\nc{\gmn}{G(m,m+n)}
\nc{\gmnk}{\gmn_{k}}
\nc{\cO}{{\cal O}}
\nc{\bcO}{\bar{\cO}}
\nc{\bO}{\bar{O}}
\nc{\hX}{\hat{X}}
\nc{\oQ}{\overline{Q}}
\nc{\ie}{{\it i.e.~}}
\nc{\eg}{{\it e.g.~}}

\def\nct{noncommutative }

\def\com{commutative }

\begin{document}
\global\parskip=4pt

\thispagestyle{empty}
\setcounter{page}0
\begin{flushright}
SU-ITP-01-043\\
IC/2001/58\\
{hep-th/0110291}
\end{flushright}
\vspace*{0.5in}
\begin{center}
{\Large\bf{  Coherent State Induced Star-Product \\[.3cm]
on $\rr^3_{\lambda}$ and the Fuzzy Sphere}}\\

\vskip .3in
{\bf {A.B. Hammou}} $^{1}$, {\bf{M. Lagraa}} $^{1,2}$, {\bf{M.M. Sheikh-Jabbari}} $^{3}$,

\vskip .5cm
$^{1}$ {\it The Abdus Salam International Center for Theoretical Physics \\
Strada Costiera 11, Trieste, Italy}\\
and\\
$^{2}$ {\it Laboratoire de Physique Th\'eorique\\
Universit\'e d'Oran Es-S\'enia, 31100, Alg\'erie}
and\\
$^{3}$ {\it Department of Physics, Stanford University\\
Stanford CA, 94305-4060, USA}

\end{center}

 
\vskip .50in
\begin{abstract}

Using the Hopf fibration and starting from a four dimensional
noncommutative Moyal plane, $\rr^2_{\theta}\times\rr^2_{\theta}$, we
obtain a star-product for the noncommutative (fuzzy) $\rr^3_{\lambda}$ 
defined by $[x^i,x^j]=i\lambda\epsilon_{ijk}x^k$. Furthermore, we show that 
there is
a projection function which allows us to reduce the functions on
$\rr^3_{\lambda}$ to that of the fuzzy sphere, and hence we introduce a
new star-product on the fuzzy sphere.  We will then briefly discuss how 
using our method one can extract information about the field theory on 
fuzzy sphere and $\rrlam$ from the corresponding field theories on 
$\rrTth\times\rrTth$ space.

\end{abstract}
\newpage
\begin{small}
\end{small}

\setcounter{footnote}{0}

\renewcommand{\baselinestretch}{2.0}
\baselineskip 0.65 cm
\def \fs {fuzzy sphere }

\newsection{Introduction and Preliminaries}

In the past two years, motivated by string theory \cite{SW}, the theories on \nct
Moyal plane has been extensively studies (for a review see \cite{DN}). The Moyal plane can
be defined through the functions of operator valued coordinates $\hX^i$ satisfying
\be\label{Moyal}
[\hX^i,\hX^j]=i\Theta^{ij}\ ,\ \ \ \ \ i=1,\cdots, d\ ,
\ee
where $\Theta^{ij}$ is a constant anti-symmetric tensor. We will denote such spaces by
$\rr^d_{\Theta}$. Let us restrict ourselves to the \nct spaces (not space-times) and take
$d=3$. In this case, always there exists a rotation which reduces the non-zero components of 
$\Theta_{ij}$ to two\footnote{We would like to comment that this is not always possible for 
the compact \nct three manifolds such as \nct three torus.},
e.g. $\Theta_{12}=-\Theta_{21}=\theta$ and therefore $\rr^3_\Theta\simeq \rrTth\times\rr$. 
As the first generalization of the Moyal plane one may consider $\Theta_{ij}$ to be
linearly $X$ dependent, \ie
\be\label{R3}
[\hX^i,\hX^j]=i\lambda\epsilon^{ijk}\hX^k\ ,\ \ \ \ \ i=1,2,3\ .
\ee
We will denote this space by the $\rrlam$. It has been shown that these spaces can also
arise within string theory \cite{{Schomerus},{Ho},{Hash},{Oz}}. The 
Eq.(\ref{R3}) 
resembles the $su(2)$ algebra whose
generators are ${\hX^i\over\lambda}$. In fact in general $\hX^i$'s are in 
reducible representations of $su(2)$. The irreducible representations of that 
algebra given
by $(2J+1)\times (2J+1)$ hermitian matrices, will reduce the $\rrlam$ to 
what is called the fuzzy
sphere, $S^2_{\lambda, J}$ 
\cite{{Madore},{GKP},{Grosse},{Bala},{Peter1},{Vaidya}} \footnote{ Recently 
there have been a complete review over the field \cite{thesis}.}. 
In other words the fuzzy sphere   
$S^2_{\lambda, J}$ is determined by the algebra (\ref{R3}) subjected to
\be\label{R3J}
\sum_{i=1}^3 \hX_i^2=\lambda^2 J(J+1)\ ,
\ee
i.e. the radius of the fuzzy sphere is given by $\lambda\sqrt{J(J+1)}$. 
Therefore the full
$\rrlam$ can be
obtained when we consider the set of fuzzy spheres with all possible radii: 
$\rrlam=\sum_{J=0}^{\infty} S^2_{\lambda, J}$. We should warn the reader that 
$\rrlam$ in the $\lambda\to 0$ limit will not reduce to $\rr^3$ \cite{GKP}.
This will be seen more explicitly in section 2 (and in particular subsection 
2.3).
It has been shown that the fuzzy sphere
in certain limits can be reduced to  the \com sphere $S^2$ and also the Moyal plane
$\rrTth$ \cite{Madore-Chu}.

In order to formulate physics on \nct spaces, one should be able to pass to the language of
fields (functions) instead of operators, where the algebra of operators is translated to the
algebra of functions on a proper space, though with a product different from the usual
product of functions. This product is usually called star-product. In fact there exists a
one-to-one correspondence (called Weyl correspondence) between the operators and the
functions \cite{{Stern},{Weyl}}: given an operator $\cO_f$
\be
\cO_f=\int dk\ {\rm e}^{ik\cdot\hX} \tilde{f}(k)\ ,  
\ee
the corresponding function is\footnote{We only consider the functions which admit the 
Fourier expansion.}
\be
f(x)=\int dk\ {\rm e}^{ik\cdot x} \tilde{f}(k)\ .
\ee
Then, the algebra of $\hX$ will induce a star product on functions:
\bea
\cO_f\cdot \cO_g&=&
\int dkdp\  {\rm e}^{ik\cdot\hX} {\rm e}^{ip\cdot\hX} \tilde{f}(k)\tilde{g}(p)\cr   
&=& \int dkdp\  {\rm e}^{i(k+p)\cdot\hX-{i\over 2}k_ip_j[\hX^i,\hX^j]+\cdots} 
\tilde{f}(k)\tilde{g}(p)\ .  
\eea
For the case of Moyal plane, the above Hausdorff expansion terminates and then we obtain the
so-called Moyal star product
\be\label{starp}
\cO_f\cdot \cO_g \longleftrightarrow f\star g= {\rm e}^{i\Theta^{ij}{\partial\over \partial
x_i}{\partial\over \partial y_j}}\ f(x)g(y)|_{x=y}\ .
\ee
It is easy to check that $x^i\star x^j=x^i x^j+{i\over 2}\Theta^{ij}$ and hence 
$\{x^i, x^j\}=x^i\star x^j-x^j\star x^i=i\Theta^{ij}$.

In running the above Weyl-Moyal machinery we have implicitly {\it assumed} 
that 
$$ 
{\rm e}^{ik\cdot\hX}\longleftrightarrow {\rm e}^{ik\cdot x}\ \ {\rm or\ equivalently}\ \ 
{\rm e}^{ik\cdot x}=({\rm e\star})^{ik\cdot x}=1+ik\cdot x-{1\over 
2!}(k\cdot x)\star(k\cdot
x)+\cdots\ .
$$
However, to obtain the algebra  (\ref{Moyal}) we do not necessarily need to impose
this condition (which is in fact a special way of ordering, the Weyl ordering). More
explicitly one can define infinitely many star-products all resulting in the same
algebra. For example, in Eq. (\ref{starp}) if we add a general symmetric matrix to
$\Theta^{ij}$ we will find the same algebra as before. This is equivalent to taking
\be\label{generalsp}
x^i\star x^j=x^i x^j+{i\over 2}\Theta^{ij}+ A^{ij}, 
\ee 
where $A^{ij}$'s are constants and $A^{ij}=A^{ji}$.
In fact, the above generalized star
products correspond to different ways of ordering in the operator language.
For the $\rrTth$ case if we choose $A^{ij}={L\over 2}\delta^{ij}$ (this is always
possible with a proper rotation), we will obtain a new star product in 
which
\be\label{weight}
({\rm e\star})^{ik\cdot x}={\rm e}^{ik\cdot x}{\rm e}^{-Lk^2/4}\ .
\ee
More precisely, different star products of (\ref{weight}) resulting from 
different orderings can be related by introducing the proper weight 
functions into the "Tr" over the algebra (which in the Moyal case is simply
integral over the whole space).
Physically this means that instead of the usual simple waves we are expanding our fields
in terms of wave packets of the width ${\sqrt{L/2}}$. It is easy to show that the
above star product will lead exactly to the same field theory results as the Moyal star 
product. More explicitly different \nct versions of a given field theory (corresponding to 
different star-products resulting from different $A^{ij}$'s) are all related by a field 
redefinition \footnote{We are grateful to L. Susskind for a discussion on 
this point.}.

Another natural way of ordering arises if instead of Fourier expansion we use the Laurent
expansion of the functions and the corresponding harmonic oscillator basis \cite{Harvey}.
Let us consider $\rrTth$ and define
\be\label{scale} 
z={x^1+ix^2\over \sqrt{2\theta}}\ ,
\ee
then $[z,\bar z]_{\star}=1$. Any function $f(x^1, x^2)$ can be expanded as
\be
f(z,\bar z)=\sum_{n,m} f_{mn} {\bar z}^m z^n\ . 
\ee
Now replacing $z$ and $\bar z$ by harmonic oscillator creation and annihilation operators 
$a$ and $a^\dagger$ we will obtain the corresponding operator which is "normal ordered".
We will show in section 2 that this normal ordering yields in the following star product:
\be\label{z*z}
z\star \bar z=z\bar z+1\ ,\ \ \ \ \
\bar z\star z =z\bar z\ ,
\ee
which exactly corresponds to Eq. (\ref{generalsp}) with $A^{ij}={\theta\over
2}\delta^{ij}$.

In order to study field theories on the $\rrlam$ we need to build the corresponding star
product. Along the above arguments, depending on the ordering we use for the operators we
will find various star products on $\rrlam$ (and similarly on $S^2_{\lambda, J}$).
If we take the Weyl ordering (i.e. imposing the condition 
$({\rm e\star})^{ik\cdot x}= {\rm e}^{ik\cdot x}$) we will end up with the 
following star
product
\be
x^i\star x^j=x^i x^j+{i\lambda\over 2}\epsilon^{ijk}x_k\ . 
\ee
However, this star product is not so convenient for doing field theory on $\rrlam$ (it is
suitable for perturbative expansions in powers of $\lambda$). In this work using the normal
ordering of operators on the Moyal plane, we obtain a new star product on $\rrlam$. To obtain 
the star product we start with a four dimensional Moyal plane, $\rrTth\times\rrTth$, 
parameterized by $z_1, z_2$ and choose
the star product induced by the normal ordering. Recalling the Hopf firbration for
$\rrlam$ \cite{thesis}, we show that the algebra of operators on $\rrlam$ (for 
$\lambda=\theta$) 
is equivalent to a sub-algebra of the functions on $\rrTth\times\rrTth$ which are invariant 
under $z_1\to {\rm e}^{i\alpha} z_1$ and $z_2\to {\rm e}^{i\alpha} z_2$, or equivalently
$\rrlam\simeq (\rrTth\times\rrTth)/S^1$. In this way one can read off the form 
of the star
product in $\rrlam$ induced by the star product on the four dimensional Moyal 
plane.
In other words there exists a dictionary which allows us to translate $\rrlam$, the algebra 
of functions and hence the field theory on that,
into that of $\rrTth\times\rrTth$. As we discussed, the
representations of the $\rrlam$ can be understood as a sum of irreducible representations
on fuzzy spheres with different radii. We show that there is a projection operator, $P_J$,
which projects the functions on  $\rrlam$ on the $S^2_{\lambda, J}$. Hence we can extend our
dictionary to translate the field theories on the \fs in four dimensional field theories on
the Moyal plane.

The paper is organized as follows. 
In section 2, we first review the harmonic oscillator basis and coherent states and then
use this basis to extract a new star product on the Moyal plane. 
We use this star product to read off the induced star product on the $\rrlam$. 
We also show how the operators (and functions) and in particular the derivative 
operators on $\rrlam$ are related to the four 
dimensional  operators (and functions).
In section 3, we introduce the projection operator $P_J$ which enables us to single out an
irreducible $(2J+1)\times (2J+1)$ dimensional representation out of the algebra of functions
on the $\rrlam$. We have moved some other useful identities involving 
$P_J$ to the Appendices.
In section 4,   we discuss how using our dictionary the field theories on 
$\rrlam$ and the \fs
can be studied through field theories on the four dimensional Moyal plane.
The last section contains our conclusions and discussions.

\setcounter{footnote}{0}

\newsection{Star Product On $\rr^{3}_{\lambda}$}

In this section, first we will review and generalize the star product 
deduced from coherent 
states \cite{{GKP},{Peter1},{APS}} and then we will construct the star product 
on 
fuzzy 
three-dimensional vector space $\rrlam$ and its projection on the 
fuzzy spheres with given radius.

The sphere can be interpreted as the Hopf fibration
\bea
S^3 =\{ \vec{z} \in {\CC}^2;\ \bar{z} z =\rho^2 \} \ \to \ S^2
=\{ x=(x^1 ,x^2 ,x^3 )\in {\rr}^3 \}
\eea
with 
\be\label{Xi}
x^i =\frac{1}{2} \bar{z}_{\alpha}\sigma^{i}_{\alpha\beta} z_{\beta}\  ,\ \ \ i =1,2,3
\ee
where the bar denotes complex conjugation and $\sigma^i$'s are Pauli matrices.
In this 
approach fields are functions of complex variables $z_{\alpha}$, $\bar{z}_\alpha$,
$\alpha =1,2$.  The relation $\bar{z} z =\rho^2$ leads to 
$\sum_{i} (x^i)^2={x^0}^2$, with 
\be\label{X0}
x^0=\frac{1}{2} \bar{z}_{\alpha} z_{\alpha}\ .
\ee
Since $x^i$'s are invariant under the transformation 
$$
z \rightarrow e^{i \alpha} \, z \quad , \quad \bar{z} \rightarrow
e^{-i \alpha} \, \bar{z} \, , 
$$
the above Hopf fibration can be viewed as coordinates on $S^2=CP^1\equiv S^3/U(1)$.

To get a noncommutative version of the above Hopf fibration it is enough
to make the coordinates $z_{\alpha}$ and $\bar{z}_{\alpha}$ 
be noncommutative: $[z_\alpha,\bar{z}_{\beta}]_{\star}=\delta_{\alpha\beta}$ (with the 
$\star$-product defined in Eq.(\ref{z*z})). Then the corresponding operator language is 
obtained by  replacing coordinates $z_{\alpha}$ and $\bar{z}_{\alpha}$ 
with  the creation and annihilation operators of a two dimensional harmonic oscillator 
$a_{\alpha}$ and $\adg_{\alpha}$
$$
[a_{\alpha},a_{\beta}^{\dagger}]=\delta_{\alpha\beta}.
$$
We note that the coordinates $z_{\alpha}$ are scaled so that they are dimensionless (as in 
Eq.(\ref{scale})) and hence $\theta$ is scaled to one. However, $\theta$ can always be 
reintroduced on a dimensional analysis.

Given 
\be
\hat{X}^i={1\over 2} a^\dagger_\alpha \sigma^i_{\alpha\beta} a_\beta
\ee
it is straightforward to show that
$$
[\hat{X}^i,\hat{X}^j]=i\eps^{ijk}\hat{X}^k\ ,
$$
(if we reintroduce $\theta$, the above will become  
$[\hat{X}^i,\hat{X}^j]=i\theta\eps^{ijk}\hat{X}^k$).
This is the key observation which relates $\rr^2_{\theta}\times\rr^2_{\theta}$ to the 
$\rrlam$ (with $\lambda=\theta$).
In this section using the above realization we obtain an explicit form of the 
star product on $\rrlam$.

\subsection{Coherent States}

Let $|n_{1},n_{2}\rangle$ represent the energy eigenstates of the 
two-dimensional harmonic oscillators whose creation and annihilation 
operators $a_{\alpha}^{\dagger}$ and $a_{\alpha}$, ($\alpha=1,2$), satisfy the 
above commutation relations. 
To any vector $\vec{z}\in {\CC}^{2}$ one can assign a coherent state 
\be\label{coh}
|z_1,z_2\rangle \equiv |\vec{z}\rangle =  
e^{-\frac{\overline{z}z}{2}}e^{z_{\alpha}a_{\alpha}^{\dagger}}|0,0\rangle
\ee
where $\bar{z}z=\bar{z}_{\alpha}z_{\alpha}$. The coherent states 
$|\vec{z}\rangle$ are normalized $\langle\vec{z}|\vec{z}\rangle=1$, form 
an (overcomplete) 
basis for the Hilbert space ${\cal H}$ and are eigenstates of the annihilation 
operators $a_{\alpha}|\vec{z}\rangle=z_{\alpha}|\vec{z}\rangle$. They are 
not orthonormal but satisfy
\begin{eqnarray}
\langle\vec{\eta}|\vec{z}\rangle=e^{-\frac{\bar{\eta}\eta}{2}-
\frac{\bar{z}z}{2}+\bar{\eta}z}.
\end{eqnarray}
The completeness relation reads
\begin{eqnarray}
\int d\mu(\bar{z},z)|\vec{z}\rangle\langle\vec{z}|=1\ ,
\end{eqnarray}
where $d\mu(\bar{z},z)=\frac{1}{\pi^{2}}d\bar{z}_{1}dz_{1}
d\bar{z}_{2}dz_{2}$ 
is the measure on the two-dimensional complex plane $\CC^{2}$. 

To any operator $\hat{f}$ belonging to the algebra $\hat{\cal A}_4$ generated 
by the creation and annihilation operators, we can associate a function 
$f(\vec{z},\vec{\bar{z}})$ belonging to the algebra of functions 
on $\rr^2_\theta\times\rr^2_\theta$ denoted by ${\cal A}_4$ and generated by $z_{\alpha}$ and 
$\bar{z}_{\alpha}$ as 
\begin{eqnarray}
\langle \vec{z}|\hat{f}|\vec{z}\rangle=f(\vec{z},\vec{\bar{z}}).
\end{eqnarray}
Then the product of operators corresponds to an associative star product 
of the corresponding functions as 

\begin{eqnarray}
(f\star g)(\vec{z},\vec{\bar{z}})=\langle \vec{z}
|\hat{f}\hat{g}|\vec{z}\rangle= \int d\mu (\bar{\eta},\eta)\langle 
\vec{z}|\hat{f}|\vec{\eta}\rangle\langle\vec{\eta}|\hat{g}|\vec{z}\rangle.
\label{star}
\end{eqnarray}

To get the explicit form of the star product (\ref{star}),  following 
\cite{APS} we introduce the translation operators

\begin{eqnarray}
e^{-z_{\alpha}\frac{\partial}{\partial \eta_{\alpha}}+{\eta}_{\alpha}
\frac{\partial}{\partial z_{\alpha}}} f(\vec{z},\vec{\bar{z}})
=\frac{\langle \vec{z}|\hat{f}|\vec{\eta}\rangle}{\langle \vec{z}|
\vec{\eta}\rangle}= 
:e^{(\eta_{\alpha}-z_{\alpha})\frac{\partial}{\partial z_{\alpha}}}:
f(\vec{z},\vec{\overline{z}})
\label{trans}
\end{eqnarray}
where : : means that the derivatives are ordered to the right in each term in 
the Taylor expansion of the exponential. Substituting (\ref{trans}) into 
(\ref{star}) and performing the integration we obtain
\begin{eqnarray}
(f\star g)(\vec{z},\vec{\bar{z}})=
f(\vec{z},\vec{\bar{z}})~~~{\rm exp}{\frac{\ld{\partial}}{\partial z_{\alpha}}
\frac{\rd\partial}{\partial \bar{z}_{\alpha}}}~~~g(\vec{z},\vec{\bar{z}}).
\label{star4}
\end{eqnarray}
This is a new star product which is resulting from the normal ordering in 
the operator 
language. We note that this $\star$-product is different from the Moyal 
star product \cite{APS}.

Besides the coherent states $|z_\alpha\rangle$ we have the usual two dimensional harmonic 
oscillator basis $|n_1,n_2\rangle$
$$
|n_1,n_2\rangle=\frac{(a_{1}^{\dagger})^{n_1}}{\sqrt{(n_1)!}}\frac{(a_{2}^{\dagger})^{n_2}}
{\sqrt{(n_2)!}}|0,0\rangle\ .
$$ 
However, it turns out that for our purpose (reduction of the four dimensional algebra to that 
of $\rrlam$) it is more convenient to use the Schwinger basis:
\begin{eqnarray}\label{Schwinger}
|j,m\rangle=\frac{(a_{1}^{\dagger})^{j+m}}{\sqrt{(j+m)!}}\frac{(a_{2}^{\dagger})^{j-m}}
{\sqrt{(j-m)!}}|0,0\rangle
\end{eqnarray}
where $j=0,\frac{1}{2},1,....\infty$ and $m$ runs by integer steps over the 
range $-j\leq m \leq j$. The coherent state can be expanded in the  $|j,m\rangle$ 
basis 
\begin{eqnarray}\label{zjm}
|\vec{z}\rangle=\sum_{j=0}^{\infty}\frac{e^{-\frac{\overline{z}z}{2}}}
{\sqrt{(2j)!}}\sum_{m=-j}^{m=j}z_{1}^{j+m}z_{2}^{j-m}\sqrt{C^{2j}_{j+m}}\,\,
|j,m\rangle
\end{eqnarray}
with $C^{2j}_{j+m}=\frac{(2j)!}{(j+m)!(j-m)!}$.

Now we consider the sub-algebra $\hat{\cal A}_{3} \subset 
\hat{\cal A}_4$ generated by $\hat{X}^{i}$, whose corresponding 
sub-algebra of functions ${\cal A}_3 \subset {\cal A}_4$ 
is generated by $x^{i}=\frac{1}{2}\bar{z}_{\alpha}\sigma^{i}_{\alpha\beta}
z_{\beta}$. Noting that $\hat{X}^0$ 
($\hat{X}^0(\hat{X}^{0}+1)=\sum_i {(\hat{X}^i)}^2$) commutes with 
all $\hat{X}^i$'s (i.e. $[\hat{X}^0, \hat{f}(\hat{X}^i)]=0$ for any 
function $\hat{f}$), 
one can define the $\hat{\cal A}_{3}$ algebra as the sub-algebra of $\hat{\cal A}_{4}$ whose 
elements are commuting with $\hat{X}^0$, i.e.
\be\label{A3def}
\forall \hat f(a^\dagger, a)\in \hat{\cal A}_{4}\; ,\; [{\hat{X}}^0, \hat{f}]=0 
\Longrightarrow\hat{f}\in \hat{\cal A}_{3}\ . 
\ee
At the level of functions the commutator with ${\hat{X}^0}$ corresponds to the
derivative operator ${\cal L}_0$:
\be\label{calL0}
i[{\hat{X}^0}, \hat{f}]\;\longleftrightarrow\; {\cal L}_0 f\equiv{i\over 
2}(
\bar z_\alpha \bar\partial_\alpha- z_\alpha \partial_\alpha)f(z,\bar z)\ .
\ee
Therefore, the elements of the algebra ${\cal A}_{3}$  are functions of 
$z_\alpha$ and $\bar 
z_\alpha$ subjected to ${\cal L}_0 f(z,\bar z)=0$.
We would like to stress that the operator ${\cal L}_0$ is in fact a derivative operator with 
respect to the star product (\ref{star4}):  
\be
{\cal L}_0 (f\star g)= ({\cal L}_0 f)\star g +f\star ({\cal L}_0 g)\ ,
\ee
and hence the sub-algebra ${\cal A}_{3}$ is closed under the star product 
(\ref{star4}), i.e. ,
$f,g\in {\cal A}_{3}$ then $f\star g\in {\cal A}_{3}$. 
Using this property and the fact that all the elements of ${\cal A}_{3}$ can be represented as 
functions of $x^i$ and $x^0$,  one can rewrite the star product of 
(\ref{star4}) in terms of 
$x^i$'s 
and their derivatives. To start with, we recall that
\bea\label{di}
\frac{\partial}{\partial z_{\alpha}}
&=&\frac{1}{2}\bar{z}_{\beta}\sigma^{i}_{\beta\alpha} \frac{\partial}{\partial 
x^{i}}\ , \nonumber\\
\frac{\partial}{\partial \bar z_{\alpha}} 
&=&\frac{1}{2}\frac{\partial}{\partial x^i}
\sigma^{i}_{\alpha\beta} {z}_{\beta}\ .
\eea
Note that the above expressions are only true when derivatives are acting 
on the functions in ${\cal A}_3$.
Now using the relation 
$$
\bar{z}_{\beta}\sigma^{i}_{\beta\alpha}\sigma^{j}_{\alpha\rho}z_{\rho}=
2(\delta^{ij}x^{0}+i\epsilon^{ijk}x^{k})\ ,
$$
we obtain the desired  star product in terms of the coordinates $x^i$ 
\begin{eqnarray}
(f\star g)(\vec{x})=
{\rm exp}\left[\frac{1}{2}(\delta^{ij} x^{0}+i\epsilon^{ijk} x^{k})
{\partial \over \partial u^i}{\partial \over \partial v^j}\right]
f(\vec{u}) g(\vec{v})\Big|_{u=v=x},
\label{star3}
\end{eqnarray}
for any functions $f, g \in {\cal A}_3$. Note that the 
exponential in the expression of the star product (\ref{star3}) should be 
understood by its Taylor expansion.
We would like to stress that our new star product (\ref{star3}), similar 
to that of Eq.(\ref{star4}) is associative.
{}From Eq.(\ref{star3}), it follows that
\begin{eqnarray}
x^{i}\star x^{j}&=&x^{i}x^{j}+\frac{1}{2}(\delta^{ij}x^{0}+i
\epsilon^{ijk}x^{k})
\label{ij}\\
x^{0}\star x^{i}&=&x^{0}x^{i}+\frac{1}{2}x^{i}
\label{0i}\\
x^{0}\star x^{0}&=&x^{0}(x^{0}+\frac{1}{2})\ , 
\label{00}
\end{eqnarray}
where we have used the fact that $x^{0}=\sqrt{x^{i}x^{i}}$ and hence 
$\frac{\partial x^{0}}{\partial x^{i}}=\frac{x^{i}}{x^{0}}$. 
To show the relations (\ref{0i}) and (\ref{00}) one can use the following
equivalent definition
$$
(f\star g)(x^i,x^0)=
f\left(x^i+\frac{1}{2}(\delta^{ij}x^0+i\epsilon^{ijk}x^k)
\frac{\partial}{\partial y^j},
x^0+\frac{1}{2}x^i\frac{\partial}{\partial y^i}\right) g(y^i,y^0)\Big|_{y=x}\ .
$$
The above is valid for any function which admit Taylor expansion.
Using (\ref{ij}) it is easy to check that 
\begin{eqnarray}
[x^{i},x^{j}]_{\star}=i\epsilon^{ijk}x^{k}\ .
\label{com}
\end{eqnarray}
Furthermore, 
$$
[x^0,x^i]_{\star}=0\ , 
$$
$$
\delta_{ij}x^{i}\star x^{j}=\vec{x}\star \vec{x}=x^{0}\star(x^{0}+1).
$$
The equation (\ref{com}) shows that the algebra ${\cal A}_3$ equipped 
with the star product (\ref{star3}) can be viewed as an algebra of functions 
on the $\rrlam$ endowed with a Euclidean metric $\delta_{ij}$. It is easy 
to see that the 
star product (\ref{star3}) is invariant under the classical $SO(3)$ group.
This $SO(3)$ symmetry of $\rrlam$ is the residual symmetry of the Poisson 
structure in $\rr^2_{\theta}\times\rr^2_{\theta}$ which is $Usp(1)\times
Usp(1)$ moded out by the $U(1)$ factor.
We would like to note that, as it is clear from our construction,  there are
two ways for computing the  star product of any functions of $x^i$'s, to use definition 
(\ref{star3}) or, to consider function as functions of  $z_{\alpha}$ and $\bar{z}_{\alpha}$ 
and use (\ref{star4}); and of course the result would be the same.

It is worth noting that if instead of the four dimensional star product of 
(\ref{star4}) one 
uses the usual Moyal star product, the reduction to a three dimensional 
star product which is 
expressible only in terms of $x^i$'s, unlike (\ref{star3}), will not have a simple form.

\subsection{The Measure}

To formulate field theory on $\rrlam$ we need to 
find the corresponding measure which should depend only on $x^i$'s and should
be related to the four dimensional measure on $\rr^2_\theta\times\rr^2_\theta$ given by
\be\label{r4measure}
d\mu (z,\bar{z})=\frac{1}{\pi^2} d\bar{z}_1 dz_1 d\bar{z}_2 dz_2.
\ee
Let us write $z_\alpha$ in a more convenient basis
$$
z_1=R \cos{\theta_3} e^{i\theta_1},\\
z_2=R \sin{\theta_3} e^{i\theta_2},
$$
with $0\leq \theta_3 \leq \frac{\pi}{2}$ and 
$0\leq \theta_{\alpha}\leq  \pi$. In this coordinate system the measure 
takes the form 
$$
d\mu (z,\bar{z})=\frac{1}{(2\pi)^2} R^3 dR \sin{2\theta_3} 
d{2\theta_3} d\theta_1
d\theta_2\,.
$$
One can easily check that $x^{i}$ and $x^{0}$ (and hence any function of them) 
depend only on $R$, 
$\theta=2\theta_3$ and $\phi=\theta_{2}-\theta_{1}$. Therefore, without any 
loss of generality we can write
\be
\int d\mu(z,\bar{z}) f(x^i,x^0)=\frac{1}{2\pi}\int_{0}^{\infty} R^3 dR 
\int_{0}^{\pi} \sin{\theta} d\theta \int_{0}^{2\pi} d\phi f(x^i,x^0),
\ee
where we have performed the integration over $\theta_{2}+\theta_{1}$ which
gives a factor of $2\pi$. Remembering that $R^2=\bar{z}z=2x^0$, $x^i$'s in this basis will be of the form
$$
x^1=x^{0}\sin{\theta} \cos{\phi},\,\,
x^2=x^{0}\sin{\theta} \sin{\phi},\,\,
x^3=x^{0}\cos{\theta},
$$
with $\sum (x^i)^2=(x^0)^2$, which is the spherical coordinate basis.
Finally the measure can be expressed as
\be\label{r3measure}
\int d\mu(z,\bar{z}) f(x^i,x^0)=\frac{1}{\pi}\int_{0}^{\infty} x^0 dx^0 
\int_{0}^{\pi} \sin{\theta} d\theta \int_{0}^{2\pi} d\phi f(x^i,x^0)=
\int {d^3x\over \pi x^0}\  f(x^i, x^0)\ .  
\ee
That is, the measure on $\rrlam$ differs from the usual $d^3x$ in a factor of 
${1\over x^0}$. This in particular implies that the radial part of 
the line element for $\rrlam$, which is the set of all possible fuzzy spheres 
with different radii, is different from the usual $\rr^3$.  
However,  the measure on the fuzzy sphere which is the angular part of the 
measure of $\rrlam$ remains the same as usual.

\newsection{Reduction to The Fuzzy Sphere}

The fuzzy sphere, $S^2_{\lambda, J}$, following the discussions in the introduction, is 
defined as $(2J+1)\times (2J+1)$ irreducible representation of the $su(2)$ algebra of 
$\rrlam$. In this section, introducing a proper projection operator, $\hat{P}_{J}$, we 
show how the star product of functions on $\rrlam$ can also be used as the 
star product on fuzzy sphere $S^2_{\lambda, J}$.

\subsection{The Projection Operator, $\hat{P}_{J}$}

Let $\hat{\cal A}_{J}$ denote the algebra of operators on the fuzzy sphere of radius $J$, 
$S^2_{\lambda, J}$ (defined through Eqs. (\ref{R3}), (\ref{R3J})).
An arbitrary element of $\hat{\cal A}_{J}$, $\hat{f}_{J}$, can be expanded in the 
Schwinger basis for fixed $J$:
\be
\hat{f}_J=\sum_{m,m'=-J}^{J} f^J_{m,m'} |J,m\rangle \langle J,m'|,
\ee
on the other hand the operator $\hat{f}\in \hat{\cal A}_{3}$ can be written as
\footnote{Note that an arbitrary element of $\hat{\cal A}_{4}$ can be expanded as
$\hat{f}=\sum_{j,j',m, m'}f^{jj'}_{m,m'} |j,m\rangle \langle j',m'|$, while,  
elements of $\hat{\cal A}_{3}$ are of the form 
$\hat{f}=\sum_{j,m, m'}f^{j}_{m,m'} |j,m\rangle \langle j,m'|$. The latter follows 
directly from the definition of the 
$\hat{\cal A}_{3}$ (Eq. (\ref{A3def})).}
\be
\hat{f}=\sum_{j=0}^{\infty} \hat{f}_{j}\ . 
\ee
Therefore to reduce the algebra $\hat{\cal A}_{3}$ to that of $S^2_{\lambda, J}$, 
$\hat{\cal A}_{J}$, it is enough to project it as
\be
\hat{f}_{J}=\hat{P}^{\dagger}_{J} \hat{f}\hat{P}_{J}\ , 
\ee
where
\be\label{PJ}
\hat{P}_J=\sum_{m=-J}^{J} |J,m\rangle \langle J,m|\ .
\ee
It is easy to check that $\hat{P}_{J}$ is a (rank $2J+1$) projection 
operator, i.e. 
\be\label{projection}
\hat{P}_{J}^{\dagger}=\hat{P}_{J},~~~\hat{P}_{J}\hat{P}_{K}=\delta_{JK}\hat{P}_{K},
~~\sum_{J=0}^{\infty}\hat{P}_{J}=1\ .
\ee
$\hat{P}_J$ can also be studied as an operator in $\hat{\cal A}_{4}$. Noting the 
definition of the $|j,m\rangle$ basis (Eq.(\ref{Schwinger})), it is straightforward to 
check that
\begin{eqnarray}\label{aPaP}
a_{\alpha}\hat{P}_{J}=\hat{P}_{J-\frac{1}{2}}a_{\alpha},~~~
a_{\alpha}^{\dagger}\hat{P}_{J}=\hat{P}_{J+\frac{1}{2}}a_{\alpha}^{\dagger}
\end{eqnarray}
and therefore 
\begin{eqnarray}\label{aaP}
[a_{\beta}^{\dagger} a_{\alpha},\hat{P}_{J}]= 0\ . 
\end{eqnarray}
{} From (\ref{aPaP}) we learn that the projected creation and annihilation 
operators are 
zero:
$$
a_{J\alpha}^{\dagger}=\hat{P}_{J}a_{\alpha}^{\dagger}\hat{P}_{J}=0\ , \ \ \ \ 
a_{J\alpha}=\hat{P}_{J}a_{\alpha}\hat{P}_{J}=0\ , 
$$ 
and (\ref{aaP}) results in
\be
[{\hat{X}}^i, \hat{P}_{J}]=0\ , 
\ee
or equivalently $\hat{P}_{J}$ is only a function of ${\hat{X}}^0$, and hence any function 
in $\hat{\cal A}_{3}$ commutes with $\hat{P}_{J}$. (This can be used as another 
equivalent definition for $\hat{\cal A}_{3}$.) Then, 
\be
\hat{f}\in\hat{\cal A}_{3},\  \hat{f}_{J}=\hat{P}^{\dagger}_{J} 
\hat{f}\hat{P}_{J}=
\hat{f}\hat{P}_{J}\ ,
\ee
and hence 
\be\label{AJdef}
\hat{\cal A}_{J}=\hat{P}_{J}\hat{\cal A}_{3}=\hat{\cal A}_{3}\hat{P}_{J}=
\hat{P}_{J}\hat{\cal A}_{4}\hat{P}_{J}\ .
\ee
The operator $\hat{P}_{J}$ can also be expanded in terms of the coherent states. First we 
recall Eq.(\ref{zjm}):
\be\label{zj}
|\vec{z}\rangle=\sum_{j=0}^{j=\infty}|\vec{z}\rangle_{j}\ , \ \ \ \ 
|\vec{z}\rangle_{j}=\frac{e^{-{\bar{z}z\over 
2}}}{\sqrt{(2j)!}}\sum_{m=-j}^{m=j}
z_{1}^{j+m}z_{2}^{j-m}\sqrt{C^{2j}_{j+m}}|j,m\rangle\ . 
\ee
We note that $|\vec{z}\rangle_{j}$ are orthogonal to each other (
${}_j\langle\vec{z}|\vec{z}\rangle_{k}=0$ for any $j\not = k$) but not normalized to 
one. With the definition of $|\vec{z}\rangle_{j}$'s and $\hat{P}_j$ we have
\be
\hat{P}_{j}|\vec{z}\rangle=|\vec{z}\rangle_{j}\ 
\ee
and therefore
\begin{eqnarray}\label{proj}
\hat{P}_{j}=\int d\mu(\bar{z}z)|\vec{z}\rangle_{j}{}_j\langle\vec{z}|\ .
\end{eqnarray}

The above operator relations can be written in terms of functions and the corresponding 
star products (\ref{star4}) or (\ref{star3}). However, we are interested in 
the 
explicit form of the 
function corresponding to $\hat{P}_J$. To obtain that one can use the coherent states:
\begin{eqnarray}\label{zPJz}
\langle\vec{z}|\hat{P}_{J}|\vec{z}\rangle=
\frac{1}{(2J)!}e^{-\bar{z}z}(\bar{z}z)^{2J}=\frac{1}{(2J)!}e^{-2x^{0}}
(2x^{0})^{2J}=P_{J}(x^{0})\ .
\end{eqnarray}
Then Eq. (\ref{projection}) will read as
\begin{eqnarray}
P_{J}(x^{0}) \star P_{K}(x^{0})=\delta_{JK}P_{K}(x^{0})
\end{eqnarray}
with the star product given in (\ref{star3}). 
To show this last equation we should expand $P_{J}(x^{0})$ in powers of 
$x^{0}$ and then use the following identity:
\be\label{x0PJ}
(x^{0})^{2l}\star P_{J}(x^{0})=(x^{0})^{2l} P_{J-l},\,\,\,
\ee
with $P_{J-l}=0$ for $J-l<0$. The proof is shown in the Appendix I.

\subsection{More on the Projection Operator $P_J$}

So far we have shown how using projection operator $P_J$, the algebra of functions on 
$\rr^2_{\theta}\times\rr^2_{\theta}$ and on $\rrlam$ (${\cal A}_4$ and ${\cal A}_3$
respectively) can be reduced to that of the fuzzy sphere, ${\cal A}_J$ defined by Eq. 
(\ref{AJdef}). In this subsection 
we would like to elaborate more on the projection operator $P_J$ and its properties.

Let us define $f_j$ as
\be
f_j(z,\bar{z})={}_j\langle\vec{z}|\hat{f}|\vec{z}\rangle_j=
\langle\vec{z}|{\hat{P}}_j\hat{f}\hat{P}_j|\vec{z}\rangle=P_j\star f\star P_j\ ,
\ee
for any $\hat f\in {\hat{\cal A}}_{4}$. It is clear that, by definition $f_j$ 
is a function 
in ${\cal A}_{j}$. If we start with the operators in 
${\hat{\cal A}}_{3}$ instead (i.e. $\hat f\in {\hat{\cal A}}_{3}$)
then
\be
f_j(x^i,x^0)=\langle\vec{z}|\hat{f}\hat{P}_j|\vec{z}\rangle=f\star 
P_j=P_j\star f\ ,
\ee
and $f(x^i, x^0)=\sum_{j=0}^{\infty} f_j(x^i,x^0)$. 
With the above definition we have
$$
({f} \star {g})_j=f_j\star g_j
$$
and 
$$
{f_j} \star {g_{j'}}= 0, \ \ \ \ \ j\neq j'\ .
$$
By a simple analysis one can show that the $x$-dependence of $f_j(x^i,x^0)$ is of the form:
\be\label{ftildef}
f_j(x^i,x^0)=\tilde{f}_j(\tilde{x}^i) P_j(x^0)\ ,
\ee
where $\tilde{x}^i={x^i\over x^0}$ are the angular part of $x^i$'s. In other words:
\be\label{tildeff}
\tilde{f}_j(\tilde{x}^i)=(f\star P_j)/P_j.
\nonumber
\ee

Here we will show some more identities involving $P_j$ 
(some more are gathered in Appendix I) which turns out to be useful in 
working out the field theory manipulations on the fuzzy sphere.
The generators of the algebra ${\cal A}_{J}$ are
\begin{eqnarray}
x^{i}_{J}=x^{i}\star P_{J}(x^{0})=x^i P_{J-{1\over 2}} (x^0)=J \tilde{x}^i 
P_J(x^0)\ .
\end{eqnarray}
$x^0$ projected on the $S^2_{\lambda, J}$, as we expect gives the radius $J$, i.e.
\begin{eqnarray}
x^{0}_{J}=x^{0}\star P_{J}(x^{0})=J P_{J}(x^{0})\ .
\end{eqnarray}
To evaluate the above star products either of (\ref{star4}) or 
(\ref{star3}) may be used.
In the same way one can show that:
\begin{eqnarray} 
[x^{i}_{J},x^{j}_{J}]_{\star} &=& [x^{i},x^{j}]_{\star}\star P_{J}(x^{0})
=i\epsilon^{ijk}x^{k}\star P_{J}(x^{0})=i\epsilon^{ijk}x^{k}_{J}\ ,\cr
[x^{0}_{J},x^{i}_{J}]_{\star} &=& 0\ , \cr
\vec{x}_{J}\star \vec{x}_{J} &=& \delta_{ij}x^{i}_{J}\star x^{j}_{J}= 
J(J+1)P_{j}(x^{0})\ .
\end{eqnarray}
Hence, $x^{i}_{J}$ can be viewed as coordinates of a sphere of 
a given radius $J$ embedded into the fuzzy space $\rrlam$. Each sphere is 
described 
by the algebra of functions ${\cal A}_{J}\subset {\cal A}_{{3}}$ generated by 
$x^{i}_{J}$. The spheres of different radii $0\leq J <\infty$ fill the whole $\rrlam$ 
or in terms of algebras $\oplus_{J=0}^{\infty}
{\cal A}_{J}={\cal A}_{{3}}$. Finally we would like to present an important identity, 
the proof of which is shown in the Appendix:
\be
f(\tilde{x}^i) \star x^0 = f(\tilde{x}^i) x^0\ ,
\ee
where $\tilde{x}^i={x^i\over x^0}$, and therefore any function of the angular coordinates 
$\tilde{x}^i$ commutes with any function which only has a radial 
dependence. This is in 
fact what one intuitively expects as, the radial coordinate $x^0$ labels 
different 
representations of the $su(2)$ algebra.

\newsection{Field Theory on $\rr^3_{\lambda}$ and The Fuzzy Sphere}

In previous sections we have given the necessary mathematical tools 
to 
construct the 
$\rrlam$ and the $S^2_{\lambda, J}$ algebra of the operators, the algebra of 
functions and the star product on them, in terms of the four dimensional Moyal-plane, 
$\rr^2_{\theta}\times\rr^2_{\theta}$. In addition, introducing the projection 
operator, $P_J$, we discussed how the fuzzy sphere algebra is resulting from that 
of $\rrlam$.
In this section as an application  of our mathematical construction we show how the 
action of field theories on $\rrlam$ and the \fs are induced from the corresponding 
actions on the Moyal plane. And hence we can deduce field theoretical information on 
$\rrlam$ and $S^2_{\lambda, J}$ from the more familiar and simpler case of the 
Moyal plane. However for reducing the actions, besides what we have already introduced
one should know what are the derivative operators on $\rrlam$ in terms of the 
derivative operators on $\rr^2_{\theta}\times\rr^2_{\theta}$. These derivatives are 
needed for writing the kinetic terms of actions. In this section we show 
that the derivative along the radial coordinate of $\rrlam$, as we expect, is 
a discrete one. 

\subsection{Derivative Operators}

The derivative operators are  generally operators which satisfy the Leibniz rule with 
respect to star product:
$${\cal D}(f\star g)={\cal D}f \star g +
f\star {\cal D} g\ .
$$
In the Moyal plane, where the noncommutativity parameter is a constant, the usual 
$\partial_{\alpha}$ and $\bar{\partial}_{\alpha}$ are proper derivative operators. 
However, it is easy to see that the 
usual $\partial_i={\partial\over \partial x^i}$ are not good derivatives with respect 
to the star product of (\ref{star3}). It is clear that, in the operator 
language, any 
operator which acts as a commutator will satisfy the Leibniz rule. (In the Moyal 
plane i.e.,  $[a_{\alpha}, {\hat \phi}]\longleftrightarrow 
\bar{\partial}_{\alpha}\phi$  
and $[a^{\dagger}_{\alpha}, {\hat \phi}]\longleftrightarrow 
-{\partial}_{\alpha}\phi$.)

In section 3, we showed that ${\cal L}_0$ (given by (\ref{calL0})) is in 
fact a 
derivative 
operator with respect to the $S^1$ direction which is moded out for reducing 
$\rr^2_{\theta}\times\rr^2_{\theta}$ to $\rrlam$. 
{}From the $su(2)$ algebra of $\rrlam$, which is the algebra of angular momenta, we 
learn that $i[\hat{X}^i,\ .\ ]$ give the proper derivatives, but not all 
of them; 
yet the radial derivative is not specified.
In terms of functions:
\be\label{calLi}
i[{\hat{X}^i}, \hat{\phi}]\;\longleftrightarrow\; {\cal L}_i \phi\equiv
\langle z|i[{\hat{X}^i}, \hat{\phi}]| z \rangle =
{i\over 2}\sigma^i_{\alpha\beta}(\bar z_\alpha \bar\partial_\beta- z_\beta 
\partial_\alpha)\phi(z,\bar z)\ .
\ee
One can explicitly show that ${\cal L}_i$ do satisfy the Leibniz rule with respect to 
star products of (\ref{star4}) and (\ref{star3}). Using (\ref{di}) one can re-write 
${\cal L}_i$ in terms of $x^i$ and $\partial_i$, and of course the result 
is the usual 
angular momentum operator, i.e. ${\cal L}_i= \epsilon_{ijk}x^j 
\partial_k$.

As we discussed the radial coordinate $x^0$ can only take discrete values of 
$j=0,{1\over 2},1,\cdots$. Therefore, the derivative in this direction is expected 
to be a discrete difference; and hence it is not necessarily fulfilling the Leibniz 
rule. On the other hand we note that ${\partial\over \partial x^0}$, 
despite being a derivative, is not a hermitian operator. 
One can check that the operator
\be\label{delta}
\Delta={1\over 2}(\bar z_\alpha \bar\partial_\alpha+ 
z_\alpha\partial_\alpha)\ ,
\ee
which corresponds to $x^0{\partial\over \partial x^0}$, is the hermitian operator 
which appears in the kinetic terms of the actions.
To show that $\Delta$ is in fact acting like a discrete derivative it is enough to 
note that 
$$
\Delta P_j=2x^0(P_{j-1/2}-P_j)=(2jP_j-(2j+1)P_{j+1/2}).
$$
Hence for any arbitrary function of $x$
$$
\phi(x^i,x^0)=\sum_{j=0} \phi_j(x)=\sum_{j}\tilde\phi(\tilde x^i)\ P_j(x^0)\ ,
$$
we have
\bea\label{deltaphi}
P_J\star \Delta \phi(x)&=& P_J\star\sum\tilde\phi(\tilde x^i)\star \Delta P_j(x^0) \cr
&=& \sum\tilde\phi(\tilde x^i)\star P_J \star \Delta 
P_j(x^0)=2J(\tilde\phi_J(\tilde{x})-\tilde\phi_{J-1/2}(\tilde{x}))P_J\ ,
\eea
which clearly shows that $\Delta$ gives the expected discrete derivative.

We have introduced four hermitian operators which are the proper derivatives 
expressed in terms of $x^i,\ x^0$ and can be used instead of $\partial_{\alpha}$ and 
${\bar{\partial}}_{\alpha}$. We also note that   
\be\label{dercom}
[{\cal L}_0,{\cal L}_i]=0\ ,  
[{\cal L}_0,\Delta]=0\ ,  
[{\cal L}_i,\Delta]=0\ .  
\ee
The equation (\ref{dercom}) confirms our previous arguments on the reduction of the 
${\cal A}_4$ algebra to ${\cal A}_3$ and ${\cal A}_J$.

We close this part by the comment that for the formulation of field theories one 
should also include the time direction which is commutative with the space directions. 
The time derivative is therefore the same as the usual.

\subsection{Field Theories on $\rrlam\times \rr$}

Along our previous discussions, given a field theory on 
$\rr^2_{\theta}\times\rr^2_{\theta}\times\rr$ ($\rr$ stands for the time direction) 
the corresponding field theory on $\rrlam\times\rr$ is obtained by restricting the 
fields to be only $x$-dependent (or elements of ${\cal A}_3$ algebra). As an explicit 
example let us consider the scalar theory on $\rr^2_{\theta}\times\rr^2_{\theta}\times\rr$
\be\label{4daction}
S=\int dt\ d\mu({\bar z}, z)\ \left(
\partial_t\phi\star\partial_t\phi-
\partial_{\alpha}\phi\star {\bar{\partial}}_{\alpha}\phi+V_{\star}(\phi)\right)\ ,
\ee
where the star products are that of (\ref{star4}) and $V_{\star}$ is the 
potential term in which all the products between $\phi$'s are carried 
out with the star product of (\ref{star4}). We should remind the reader 
that with the star product (\ref{star4}), unlike the Moyal star product,
\be
\int d\mu({\bar z}, z)\ f \star g \neq  \int d\mu({\bar z}, z)\ f g \ ,
\ee  
i.e. we cannot remove star product in the quadratic terms of the action. However, 
still we have the cyclicity of the star product inside the integral, 
\be
\int d\mu({\bar z}, z)\ f_1 \star f_2 \star \cdots\star f_n =
\int d\mu({\bar z}, z)\ f_n \star f_1 \star \cdots\star f_{n-1}\ .
\ee  
And the same is also true for the star product of (\ref{star3}):
\be    
\int {d^3x\over x^0}\ f_1(x) \star f_2(x) \star \cdots\star f_n(x) =
\int {d^3 x\over x^0}\ f_n(x) \star f_1(x) \star \cdots\star f_{n-1}(x)\ .
\ee

Restricting $\phi$ to be only $x$-dependent, the potential term is replaced 
with the  same functional form of $\phi$, with the star product of 
(\ref{star3}). 
The kinetic term is more involved, because $\phi\in {\cal A}_3$ (${\cal 
L}_0\phi=0$), 
we cannot conclude that $\partial_{\alpha}\phi, 
\bar{\partial}_{\alpha}\phi \in {\cal A}_3$. 
However, with a little algebra one can show that
\bea\label{kinetic3}
\phi\star\square_4\phi=
\frac{1}{2} 
\phi\star(x^0\partial_i\partial_{i}\phi) \ .
\eea
Recalling the discussion of section 2.2, one can write the action 
(\ref{4daction}) in terms of $\rrlam\times\rr$ parameters:
\be\label{3daction}
S=\int dt \frac{d^3x}{\pi x^0} 
\left\{\partial_t\phi\star\partial_t\phi+
\frac{1}{2}\phi\star(x^0\partial_i\partial_{i}\phi)
+V_\star(\phi) \right\} \ .
\ee
We would like to comment that, since $\partial_i$ are not 
derivative operators with respect to the star product (\ref{star3}), the 
spatial part of the kinetic term should be handled with a special care.   
On the other hand if one tries to use the ``proper derivative'' (i.e.
$\Delta, {\cal L}_i$), as it is shown in Appendix II, the form of kinetic 
terms of the action are not as simple as inserting the star product into 
the commutative expressions.

\subsection{Field Theories on the Fuzzy Sphere}

In this section we would like to discuss how using our formalism one can 
study field theory on the fuzzy
sphere $S^2_{\lambda, J}$. As we have previously discussed in subsection 3.2,
any  function $f$ on $\rrlam$ can be written as
$$
f=\sum_j f_j(x^i,x^0)=\sum_j \tilde{f}_j(\tilde{x}^i) P_j(x^0)\, ,
$$
where $\tilde{x}^i={x^i\over x^0}$ are the angular part of $x^i$'s, and 
$\tilde{f}_j(\tilde{x}^i)$ is the corresponding function on $S^2_{\lambda, J}$ 
given by Eq.(\ref{tildeff}). 
It is easy to show that 
\bea
Tr \hat{P}_J &=&\sum_{j,m} \langle j,m|\hat{P}_J|j,m\rangle=\sum_m <J,m|J,m> 
\nonumber\\
&=&\int d\mu(z,\bar{z}) P_J(z,\bar{z}) \nonumber\\
&=&\int {d^3 x \over \pi x^0} P_J(x^0)=2J+1\,\,\, ,
\nonumber
\eea
which is giving the dimension of an irreducible representation of spin $J$. 
Using the above observation one can, therefore, deduce the field theory on
the fuzzy sphere starting from the one on $\rrlam$ by simply performing the 
integration over $x^0$ and dividing the result by the factor $2J+1$ for a 
sphere of radius $J$. Then as a result we have
\bea
\int {d\Omega \over 
4\pi}\ \tilde{f}_{1J}\star\tilde{f}_{2J}\star\cdots\star
\tilde{f}_{nJ}={1\over (2J+1)} \int {d^3 x \over \pi x^0} 
(f_1\star f_2\star\cdots\star f_n)\star P_J 
\eea
where the star product of the functions defined on $S^2_{\lambda, J}$ 
is defined in terms of the star product (\ref{star3}) of the 
corresponding 
functions on 
$\rrlam$ through equation (\ref{tildeff}).
%
%
Now the action for a scalar field, for example, on the fuzzy sphere of a given
radius $J$ will be expressed in $\rrlam$ as

\bea
S_J={1 \over (2J+1)} \int dt\ {d^3 x \over \pi x^0} 
(\partial_t \phi\star\partial_t \phi-
{\cal L}_i\phi\star{\cal L}_i\phi+V_\star(\phi))\star P_J \,\, ,
\label{actionS}
\eea 
and as we have already explained, performing the integration over the $x^0$
part will result in 
\be
S_J=\int dt\ {d\Omega \over 4\pi} (\partial_t \tilde{\phi}_J 
\star\partial_t \tilde{\phi}_J-
{\cal L}_i\tilde{\phi}_J\star{\cal L}_i
\tilde{\phi}_J+V_\star(\tilde{\phi}_J))\,\, .
\ee

Noting the fact that any $(2J+1)\times (2J+1)$ hermitian matrix can be
expanded in terms of spherical harmonic, $Y_{lm}\ \ l\leq 2J$, 
instead of working  with matrices, spherical
harmonics are usually used for field theory manipulations on the fuzzy
sphere. So, to complete our dictionary for the fuzzy sphere, we would like
to show how the $Y_{lm}$'s can be expressed in terms of $x$ coordinates.
This will automatically lead to the proper star product between the
$Y_{lm}$'s. The classical spherical harmonics,
are a set of orthonormal functions obeying
\bea
{\cal L}_i^2 \,\,Y_{l,m}&=&l(l+1) \,\,Y_{l,m}\,\, ,\nonumber\\
{\cal L}_3   \,\,Y_{l,m}&=& m \,\,Y_{l,m}\,\, .
\label{ylm}
\eea

Following \cite{GKP} we introduce the highest weight functions in ${\cal 
A}_3$ algebra
\footnote{Note that, unlike the commutative case, any function on 
$\rrlam$ can be expanded through the spherical harmonics.} 
\be
\psi_{l,l}(x) =\sum_j c_{j,l} \,\,\,x^{l}_+ \star P_{j}
\label{hw}
\ee
where $x_+=x^1+ix^2$ and  $c_l$ are normalization factors.
Then, by virtue of  (\ref{x0PJ}), for each term (of specific value of 
$j$) in the sum  $l=0,1, \cdots ,2j$  
and in addition ${\cal L}_+ \psi_{l,l}=0$ for all $(l\leq 2j)$. 
In order to reduce on the $S^2_{\lambda, J}$, one should multiply  (4.15) 
by 
$P_J$:
\be
\psi^J_{l,l}(x) =c_{J,l} \,\,\,x^{l}_+ \star P_{J}, \ \ \ l\leq 2J\ .
\label{hwJ}
\ee
Acting on $\psi^J_{l,l}$ with the operator ${\cal L}_-$ will lead 
to the functions (\ref{ylm})
\footnote{Here we will not try to compute the exact values of 
the normalization factors 
which are very important if one wants to do explicit field theory 
calculation.} 
 
\be
Y^J_{l,m}=N_{l,m} ({\cal L}_-)^{l-m}\psi^J_{l,l}
\ee
where $m$ runs by integer steps over the range $-l\leq m \leq l$. 
Any function in $S^2_{\lambda, J}$ can be expanded in terms of $Y^J_{l,m}$ 
as
\be
\Phi_J(x)=\sum_{(l,m)} a_{l,m} Y^J_{l,m}
\label{ay}
\ee
where the sum over $(l,m)$ means $l=0,\cdots , 2J; 
-l\leq m \leq l$, and $a_{l,m}$ are complex coefficients obeying
\be
a_{l,-m}=(-)^m a^*_{l,m}\,\,\, ,
\ee
which guarantees the reality condition $\Phi^*(x)=\Phi(x)$. However, we 
note that in (\ref{hwJ}) still there is  $x^0$ dependence, which can be 
removed by integration over $x^0$ or simply by dividing by $P_J$:
\be
\tilde{Y}^{J}_{l,m}=Y^J_{l,m}/P_J\,\,\, ,
\ee
and therefore, any function $\tilde{\Phi}$ on $S^2_{\lambda, J}$ can be 
expanded as in (\ref{ay})
\be
\tilde{\Phi}_J(x)=\sum_{(l,m)} a_{l,m} \tilde{Y}^J_{l,m}\,\,\, ,
\label{ayt}
\ee        
where now $\tilde{Y}^{J}_{l,m}$ depend only on the angular part of $x$ 
as it should be.

\newsection{Discussion}

In this paper we have constructed a dictionary which translates different 
descriptions of $\rrlam$ (the set of fuzzy spheres with all possible radii) 
or the fuzzy spheres into each other. The fuzzy 
sphere and $\rrlam$ may be studied through the operators which are 
functions of the coordinates $\hat{X}^i$, the generators of the $su(2)$ 
algebra. 
On the other hand always there is a corresponding algebra of functions 
with the proper star product which is equivalent to the operator algebra. 
The latter is the appropriate language for performing field theories. 
Starting from the four dimensional Moyal plane and reducing that on a circle, we 
have constructed $\rrlam$ and from there we read off the star product 
induced from the Moyal plane.
Then we discussed how the algebra of functions on the fuzzy sphere can be 
realized as different sectors of the algebra of functions on $\rrlam$ (and 
of course with the same star product). If we reintroduce $\theta$ in our 
expressions, we would obtain a factor of $\theta$ in the exponential 
factor in the Eq.(\ref{star3}). Then, it can be checked directly from the
definition of our star product (\ref{star3}) that in the $\lambda$ (or 
$\theta$) $\to 0$ limit we will find the usual product of functions.

Here we have concentrated on completing the mathematical tools. Using our 
results one can easily study the field theories on a fuzzy sphere through 
the Moyal field theories. As a result we would like to mention 
about the IR/UV mixing, which is a general feature of noncommutative 
Moyal field theories. So, using our method, we expect to be able to trace the 
same phenonemon for the fuzzy sphere 
field theories. This, in fact has been explicitly checked by using the 
spherical harmonics \cite{{Vaidya}, {Peter2}}.
There are several interesting open problems one can address here. Using
 our formulation we have a straightforward way of introducing fermions on 
the fuzzy sphere, starting from the fermions on the Moyal plane.
Furthermore, we have a simple handle on the vector gauge fields on the 
fuzzy sphere.

The other interesting question is the solitonic solutions on the fuzzy 
sphere \cite{{soliton1},{soliton2}}. In our approach, one can easily 
obtain the 
solitonic  solutions on the fuzzy sphere from the solitonic solutions on 
the Moyal plane which respect the rotational $su(2)$ symmetry.
As an explicit example we would like to note that the fuzzy sphere itself 
can be thought as a solitonic solution in the Moyal field theory; as it 
can be identified with the projector, $P_j$. We postpone a full study of 
such solutions to future works.

{\bf Acknowledgments}

We are grateful to P. Pre\v{s}najder for careful reading of the manuscript 
and his comments. A.B.Hammou would like to thank E. Gava for very helpful 
discussions. He 
would like also to thank the INFN sezione di Trieste and the High Energy 
section of SISSA-ISAS for hospitality and grants.
We would like to thank the High Energy Section of the
Abdus Salam ICTP where this work was started. The work of 
M.M. Sh-J. is supported in part by NSF grant PHY-9870115 and
in part by funds from the Stanford Institute for Theoretical Physics.

\vskip 1cm

{\bf Appendix I. Proof of some useful identities have been used in the 
paper}

The proof for Eq.(\ref{x0PJ}): 
$$
(x^0)^{2l}\star P_j(x^0)=(x^0)^{2l} P_{j-l}
$$
To show the above we should perform the star product explicitly, i.e.
$$
(x^0)^{2l}\star P_j(x^0)=
(x^0+{1 \over 2} x^i {\partial \over \partial y^i})^{2l} P_j(y^0)|_{y=x} 
\nonumber\\
=(x^0)^{2l} \sum_{n=0}^{2l} C^{2l}_n ({x^i \over 2x^0}
{\partial \over \partial y^i})^n P_j(y^0)|_{y=x}\ ,
$$
with $C^{2l}_n={(2l)!\over n!(2l-n)!}$. Since $\partial_i P_j= {x^i \over 
x^0}\partial_0 P_j$, where 
$\partial_0={\partial \over \partial x^0}$ is the derivative with respect to 
$x^0$, after straight forward calculations one can show that
\bea
(x^0)^{2l}\star P_j(x^0)&=&(x^0)^{2l} \sum_{n=0}^{2l} C^{2l}_n {1 \over 
2^n} \partial_{0}^n P_j(x^0) \nonumber \cr
&=&(x^0)^{2l} (1+{1\over 2}{\partial\over \partial x^0})^{2l} P_j\ .
\nonumber  
\eea
On the other hand it is easy to check that
$$
(1+{1\over 2}{\partial\over \partial x^0}) P_j= P_{j-1/2}\ .
$$
Therefore
$$
(x^0)^{2l}\star P_j(x^0)=\left\{\begin{array}{cc}
(x^0)^{2l} P_{j-l}(x^0)\ \ \ {\rm for}\ \  j\geq l\\
0 \ \ \ {\rm for}\ \  j<l\ .
\end{array}\right.
\nonumber
$$
Equivalently  one can prove the 
Eq.(\ref{x0PJ})  using the operator language. {}From Eq. (3.6) it follows that
$$
:(\hat{x}^0)^{2l}:\hat{P}_j= ({1\over 2})^{2l}a^{\dagger}_{\alpha_1}\cdots 
a^{\dagger}_{\alpha_{2l}}
a_{\alpha_1}\cdots a_{\alpha_{2l}} \hat{P}_j
=({1\over 2})^{2l}
a^{\dagger}_{\alpha_1}\cdots 
a^{\dagger}_{\alpha_{2l}}\hat{P}_{j-l}
a_{\alpha_1}\cdots a_{\alpha_{2l}}\ .
$$
Applying the coherent states we will find Eq.(\ref{x0PJ}). 

\vskip 1cm

{\it Some more identities}
 
\bea
x^{i}_{J}\star x^{j}_{J}&=&x^{i}\star x^{j}\star P_{J}(x^{0})=(x^{i}x^{j}+
\frac{1}{2}(\delta^{ij}x^{0}+i\epsilon^{ijk}x^{k}))\star P_{J}(x^{0})
\nonumber\\
&=& x^{i}x^{j}\frac{1}{(2J-2)!}e^{-2x^{0}}(2x^{0})^{2J-2}+\frac{1}{2}
(\delta^{ij}x^{0}+i\epsilon^{ijk}x^{k})\frac{1}{(2J-1)!}e^{-2x^{0}}(2x^{0})^
{2J-1}\nonumber\cr
&=&\left[J(J-{1\over 2}) \tilde{x}^i\tilde{x}^j + {1\over 
2}J(\delta^{ij}+i\epsilon^{ijk}\tilde{x}^k)\right] P_J\ ,
\nonumber
\end{eqnarray}
and
\begin{eqnarray}
x^{0}_{J}\star x^{0}_{J}&=&x^{0}x^{0}\frac{1}{(2J-2)!}e^{-2x^{0}}(2x^{0})^
{2J-2}+\frac{1}{2}\frac{\delta^{ij}x^{0})}{(2J-1)!}e^{-2x^{0}}(2x^{0})^{2J-1}
\nonumber\\
&=&(\frac{2J(2J-1)}{4}+\frac{J}{2})\star P_{J}=J^{2}P_{J}
\nonumber
\end{eqnarray}
from which we deduce
\begin{eqnarray}
\vec{x}_{J}\star \vec{x}_{J} = \delta_{ij}x^{i}_{J}\star x^{j}_{J} &= &   
(x^{0})^{2}\frac{1}{(2J-2)!}e^{-2x^{0}}(2x^{0})^{2J-2}+\frac{3}{2}x^{0}
\frac{1}{(2J-1)!}e^{-2x^{0}}(2x^{0})^{2J-1}\nonumber\\
&=& x^{0}_{J}\star (x^{0}_{J}+1) =J(J+1)P_{j}(x^{0})\ .
\nonumber
\end{eqnarray}

An other important relation is the star product between $x^0$ and 
$\tilde{x}^i$ 
\bea
x^0 \star \tilde{x}^i &=& 
(x^0+{1 \over 2} x^j {\partial \over \partial x^j}) \tilde{x}^i
\nonumber\\
&=& x^0 \tilde{x}^i+{1 \over 2} x^j 
({\delta^{ij} \over x^0}-{x^i x^j \over (x^0)^3})
\nonumber\\
&=& x^0 \tilde{x}^i=\tilde{x}^i \star x^0\ ,
\nonumber
\eea
which in turn results in Eq.(3.24).
Then using the above relations one can check that
$$
(x^i)^{2l}\star 
P_J(x^0)=({\tilde{x}^i})^{2l}\star({x^0})^{2l}\star P_J(x^0)=
({\tilde{x}^i})^{2l}\star (({x^0})^{2l} P_{J-l}(x^0))= 
({{x}^i})^{2l} P_{J-l}(x^0)\ .
$$

\vskip 1cm

{\bf Appendix II. More on kinetic terms}

If we write the kinetic terms of the four dimensional theory as
$$
\phi\star\square_4 \phi\ ,
$$
where
$$
\square_4=\partial_{\alpha}\bar\partial_{\alpha}\ .
$$
Then, for $\phi$'s with ${\cal L}_0\phi=0$, in terms of the three 
dimensional derivatives we have:
$$
-\phi\star(x^0\square_4\phi)=\phi\star({\cal L}_i{\cal 
L}_i+\Delta\Delta)\phi+ \phi\star\Delta\phi\ .
$$
The last term $\phi\star\Delta\phi$ is there because $\Delta$ is not a
derivative (or in other words, it is a discrete derivative in the radial
direction). On the other hand, we note that
$$
x^0\square_4\phi=\square_4(x^0\phi)-\phi-\Delta\phi\ .
$$
Consequently, we obtain
$$
- \phi\star\square_4(x^0\phi)=\phi\star({\cal L}_i{\cal 
L}_i+\Delta\Delta)\phi
+ \phi\star\phi .
$$
%
As another way of writing the kinetic term we have

\begin{eqnarray}
\int \phi\star\Delta\phi&=&\int \left({1\over 2}
\partial_{\alpha}\phi\star\bar\partial_{\alpha}\phi-\phi\star\phi\right)\nonumber\cr
&=& -{1\over 2}\int \phi\star[(\square_4+2)\phi]\nonumber\ .
\end{eqnarray}

\end{document}